\documentclass[conference]{IEEEtran}
\IEEEoverridecommandlockouts
\usepackage{cite}
\usepackage{amsmath,amssymb,amsfonts}
\usepackage{algorithmic}
\usepackage{graphicx}
\usepackage{textcomp}
\usepackage{multirow}
\usepackage{amssymb}
\usepackage{booktabs}
\usepackage{threeparttable}
\usepackage{enumitem}
\usepackage[framemethod=tikz]{mdframed}
\usepackage{tikz}
\usepackage{tipa}
\usepackage{xtab}
\usepackage{mfirstuc}
\usepackage{todonotes}
\usepackage{soul}
\usepackage{colortbl}
\usepackage{xspace}
\usepackage{xcolor}

\usepackage{url}

\usepackage{caption}
\usepackage{subcaption}
\usepackage[many]{tcolorbox}
\newtcolorbox{mybox}[1][]{
  breakable,
  title=#1,
  colback=white,
  colbacktitle=white,
  coltitle=black,
  fonttitle=\bfseries,
  bottomrule=0pt,
  toprule=0pt,
  leftrule=3pt,
  rightrule=3pt,
  titlerule=0pt,
  arc=0pt,
  outer arc=0pt,
  colframe=black,
}

\newcommand{\company}[1]{\textsf{\small #1}}
\newcommand{\q}[1]{\emph{``#1''}}

\def\BibTeX{{\rm B\kern-.05em{\sc i\kern-.025em b}\kern-.08em
    T\kern-.1667em\lower.7ex\hbox{E}\kern-.125emX}}
    
\begin{document}


\title{``Software is the easy part of Software Engineering" - Lessons and Experiences from A Large-Scale, Multi-Team Capstone Course}


\author{\IEEEauthorblockN{
Ze Shi Li, 
Nowshin Nawar Arony,
Kezia Devathasan,
Daniela Damian}
\IEEEauthorblockA{
\emph{Department of Computer Science}\\
\emph{University of Victoria, Victoria, Canada}\\
\emph{\{lize, nowshinarony, keziadevathasan, danielad\}@uvic.ca}}}


\maketitle
\thispagestyle{plain}
\pagestyle{plain}

\begin{abstract}
Capstone courses in undergraduate software engineering are a critical final milestone for students. 
These courses allow students to create a software solution and demonstrate the knowledge they accumulated in their degrees.
However, a typical capstone project team is small containing no more than 5 students and function independently from other teams.
To better reflect real-world software development and meet industry demands, we introduce in this paper our novel capstone course.
Each student was assigned to a large-scale, multi-team (i.e., company) of up to 20 students to collaboratively build software.
Students placed in a company gained first-hand experiences with respect to multi-team coordination, integration, communication, agile, and teamwork to build a microservices based project.
Furthermore, each company was required to implement plug-and-play so that their services would be compatible with another company, thereby sharing common APIs.
Through developing the product in autonomous sub-teams, the students enhanced not only their technical abilities but also their soft skills such as communication and coordination. 
More importantly, experiencing the challenges that arose from the multi-team project trained students to realize the pitfalls and advantages of organizational culture.
Among many lessons learned from this course experience, students learned the critical importance of building team trust.
We provide detailed information about our course structure, lessons learned, and propose recommendations for other universities and programs.
Our work concerns educators interested in launching similar capstone projects so that students in other institutions can reap the benefits of large-scale, multi-team development.
\end{abstract}


\section{Introduction}
A capstone course is typically adopted in universities as the "finale" course that an undergraduate student in computer science or software engineering takes to apply and demonstrate their understanding of all the skills and knowledge learned throughout their degree.
In computer science and software engineering, a capstone course will usually entail students working in teams of 3-5 students on the software they build from scratch over a semester. 
Theoretically, students will demonstrate their mastery of topics such as algorithms, architecture design, and software development. 
In many cases, students also have the freedom to select their own team, the topic, and scope of their project.

Since a capstone course is usually right before students enter industry, some universities tailor their capstones with a heavy emphasis on industry.
To this end, one method universities adopt is partnering with industry to serve as clients for capstone projects \cite{herbert2018reflections}. 
Students receive exposure to industry development through developing a software that may be used in practice and may also receive experience with client negotiation and scope. 
Moreover, some universities provide additional training such as experience with DevOps, or Agile development \cite{mason2017teaching}. 
Some computer science and software engineering programs provide courses to teach students the practices and challenges of multi-team coordination across cultural, geographic, and time-zone difference \cite{damian2006instructional, nordio2014experiment, clear2015challenges}.
Other capstone courses offered have exposed students to the experience of working in larger teams (i.e., 12 students) and exposed them to agile approaches \cite{schneider2020adopting}. 

University of Victoria has a longstanding and close relationship with organizations from industry who regularly hire co-operative education (co-op) interns and new grads from our faculty. 
Based on our industry partners' suggestions, this year we revamped our capstone course to provide students with real-world multi-team software development experience before graduation. 
Modern day software development often encompasses large-scale multi-team coordination and teamwork where software engineers work together to build a software product.
However, to the best of our knowledge, there are not yet any capstone course for software engineering that accomplishes this.

Hence, we address how we can teach students how to work on large scale and multi-team development in the classroom, so that they are better prepared for industry roles. 
In this paper, we present our novel capstone course that prescribed students with building software in large-scale, multi-team, collaborative, and self-organized groups.
77 students were split into 4 \company{companies} (i.e., large student teams) consisting of 19-20 students.
Each company built the same product, a course scheduling system for the courses in the software engineering program, with modular architecture. 
Each company also had a companion company who were required to develop a common API and share services.
The common API would allow a pair of companies to ``plug and play" different subteams with each other. 
We collected student personal reflections and teaching team reflections to derive lessons learned and make improvements for future iterations.
Not only did the students exemplify and develop their technical skills in this capstone, we identified from the lessons learned that students significantly improved their soft skills, such as communication, coordination, and knowledge management, without which it would not be possible to complete the project.
We learned the great importance of building trust and culture in a multi-team project as trust was paramount for coordinating work between \company{companies} and across sub-teams within the same \company{company}.
We conclude the paper with an extensive discussion on the many lessons learned and we hope our recommendations will prove beneficial for educators who may want to offer similar versions of our capstone.

\section{Motivation} \label{context}
This paper reports the overall design and experience of a newly structured capstone course at our university designed for students enrolled in the fourth year of their undergraduate Bachelor of Software Engineering degree (BSEng). 
As such, for most students, this course would be one of the last required courses for their university degree. 
Thus, it can safely be assumed that all students in this course were senior software engineering students. 
Additionally, most of them should have completed three or four co-op internships because to graduate from the University of Victoria with a BSEng degree, students must complete at least four co-op terms. 
Furthermore, some of the students have experience working in large companies like Amazon and Microsoft.
As a result, the students are expected to have knowledge of basic programming, software development, basic and advanced algorithms, databases, machine learning, etc. 
Most students in this course had adequate academic knowledge and real-world industry experience. 

In previous years, this capstone course was open to students from other departments like electrical and mechanical engineering. They were allowed to choose their own capstone project, and their team members consisted of three to five students.
They would work with their specific team members throughout the semester to finish the project. 
However, in recent years this course offering has been changed and is now only offered to software engineering students.
More importantly, the university collaborates closely with the local industry since the students either join these companies for co-op internships or full-time jobs after graduation. 
Due to this collaboration, the companies, on a regular basis, give reviews regarding the critical skills that the students are missing from the educational experience. 
One of the key feedback from the companies consisted of the fact that many students did not have experience working in multi-team software development settings where they required communication and coordination with other teams.
Although our students had the skills to work on small projects or problems, they lacked real-world software engineering skills such as working with different people across departments, communication, and documentation.
Prior to this capstone, students did not have the opportunity to experience large-scale software development processes. 
Therefore, the department decided to change the course structure so that the students would be exposed to as realistic conditions as possible.

\section{Related Work}
Software engineering and computer science is a discipline that is extensively industry focused, however students often lack the real world experience.
Studies have analyzed ways to improve how software engineering is taught \cite{chen2009exploring}, so students are more prepared to work in the industry.
Project based learning (PBL) is a popular method in educating students on dealing with real world problems \cite{savery2015overview}. 
Capstone courses utilize PBL methods, where students work on projects applying four-five years worth of their knowledge.
However, these courses have been part of the engineering curriculum for a while now \cite{dutson1997review}.

A number of studies have focused on improving the collaborative experience of such software engineering courses \cite{chen2011software, flowers2008improving}.
Capstone projects provide the students the opportunity to work with industry clients in an academic setting which help them improve their professional skills \cite{herbert2018reflections}. 
In addition, capstone courses provide students with a plethora of real world experiences teaching students various soft skills. 
Although students acquire technical skills during their undergraduate courses, they often lack the skills to navigate through difficult teamwork situations \cite{bastarrica2017can}. Although there is an emphasis on navigating teamwork, there is sparse research into the impacts other factors such as team size can have on the student experience.
Khakurel \emph{et al.} \cite{khakurel2020effect} described students during such capstone courses learn soft skills such a communication, collaboration, team management, and conflict resolution, which are beneficial to them when they enter the workforce after graduation. Such soft skills are especially beneficial in software engineering disciplines where team members may take on specific roles. For example, a scrum master requires certain soft skills \cite{matturro2015soft} such as effective communication and strong interpersonal skills \cite{hidayati2020hard}. In a setting where such soft skills may not be fully developed, students may experience further challenges.
In a similar thread, another key aspect of working in the industry setting includes understanding organizational culture and the ways knowledge management practices are dependent on the specific company culture \cite{de2015knowledge}. 
                           
A popular method practiced in capstone course is the agile approach.
Umphress \emph{et al.} \cite{umphress2002software} conducted an extensive study on 49 capstone projects and reported the benefit of using agile in software development. 
When introduced to agile software development, students although enjoy working with the process, they find it difficult to understand at first \cite{schneider2006agile, schneider2020adopting}.
Dikert \emph{et al.} \cite{dikert2016challenges} explains, large scale agile teams have their own challenges in implementation and integration of agile but with adequate training and management support this can be mitigated.
To address the challenges in agile, Schneider \emph{et al.} \cite{schneider2020adopting} proposed a Tribes and squads model that integrates agile in a setting with teams of 10-12 members each working on developing industry projects.
This model involved a unique collaboration between junior and senior that provided the students with relevant industry experience.  
Scrum is another widely used agile process which is integrated in capstone course \cite{schwaber2002agile}.
Christensen and Paasivaara \cite{christensen2022respond} describe an online Scrum simulation in distributed teams where the participants reported better learning outcomes in terms of communication, estimation and collaboration with industry partners.

In a study conducted on a capstone course by Paasivaara \emph{et al.} \cite{paasivaara2019collaborating}, students worked 6 months on a project utilizing scrum and various technical skills. 
The authors reported the students' learning enhances to a significant extent while working on a realistic capstone project.
Furthermore, in a study conducted by Khmelevsky \cite{khmelevsky2016ten} on reviewing 10 years of capstone projects at Okanagan College, the author described the student learning experience working with both small and large international companies.
He reports that working on capstone projects drastically improved the students learning. 
In addition, students acknowledged that communication and collaboration are key aspects to such projects which need to be emphasized. 
Bütt \emph{et al.} \cite{butt2022student} describes a new structure of capstone projects where the students sponsored their own project to gain real world experience of product business planning and risks in a startup.

The existing capstone courses do not provide adequate paradigm for large-scale multi-team cohorts where the teams need to consider inter-team integration methods utilizing scrum techniques. 
Although, Stansbury \emph{et al.} \cite{stansbury2011agile} adapted agile processes to support large
multidisciplinary teams of 12 to 16 students, our approach aims to replicate a real world software company where teams required efficient coordination to deliver the final product. 

In the following sections, we describe our newly structured Capstone course that uses scrum practices to depict a large-scale multi-team industry setting.

\begin{figure*}[h!]
 
  \centering
  \includegraphics[width=18cm]{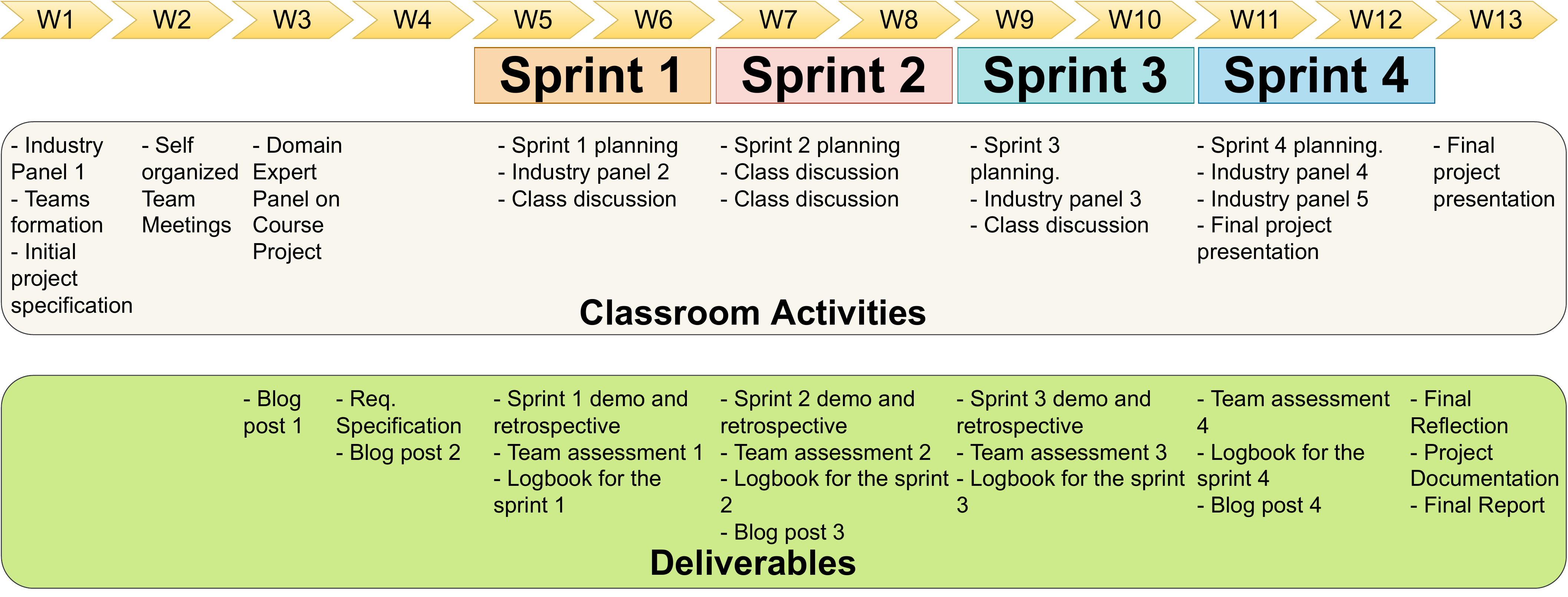}
  \caption{Course Timeline (W representing Week)}
  \label{coursetimeline}
\end{figure*}

\section{Course Setup} \label{course_setup}
The teaching team designed the structure of the course in a way that would not only test students on their university knowledge and skills but also provide them with experience working on a large-scale, multi-team project. 
This section will highlight the course structure, teaching goals, project description, project timeline and assessment criteria. 
We also provide a replication package for educators \url{https://doi.org/10.5281/zenodo.7196427}.

\subsection{Course Structure: } \label{courseStructure}
The BSEng capstone course is a mandatory 1.5 credit approximately 10 hours of work per week that the students are required to complete to graduate.
The course structure primarily focused on the concept of working in a large, multi-team project, however like any other course there was a set structure. 
The course employed scrum methodology over the course of 13 weeks.
The product development occurred over four 2-week sprints.
In total, 12 class lectures, 80 minutes each, were conducted throughout the term, which consisted of class discussions on assigned readings, industry expert panels, and one self-paced work week during the term. 
The class discussions primarily focused on topics like rules of starting a career, coordination, communication, knowledge management,  culture, and trust.

Furthermore, each team had a dedicated 5 hours and 30 minutes of lab time per week. 
The intention behind the long lab hours encompassed allocating ample time to the students where they could meet and work together, as finding time outside of class is notoriously difficult given other courses, work, commute, and personal obligations. 
Along with the final project, the students were required to submit several deliverables, each of which was part of the grading system. 
The last week was dedicated to the final presentation, where students showcased their final projects. 
Figure \ref{coursetimeline} provides an overview of a timeline for the activities and deliverables. 

\subsection{Teaching Goals: }\label{teachingGoals}
The teaching team aimed to train and provide students experience with multi-team coordination and sharing common APIs between microservices. 
Considering the emphasis on replicating real world large, multi-team industry experience, each team was referred to as a \company{company}.
Students were instructed to work in a microservice architecture, where each \company{company} would be writing Application Programming Interfaces (API) not just for their own sub-teams but for other \company{companies} as well to facilitate plug-and-play. 
Figure \ref{architecture} demonstrates the microservice architecture between two companies. 
Implementing all these factors in the course was to assess and teach the students the following:

\begin{itemize}
    \item current state of practice in software development.
    \item plan and work in a large-scale, multi-team project and coordinate with multiple sub-teams using scrum practices.
    \item develop teamwork, communication and collaboration skills.
    \item demonstrate proficiency of technical skills regarding databases, algorithms, and architecture, etc.
    \item plan project timelines with milestones and deadlines.
\end{itemize}

\subsection{Project Description:}
With the vision to replicate the multi-team industry experience, this year's project required designing and implementing a university course scheduler for our university's software engineering and computer science department using scrum practices.
Although a course scheduling program sounds straightforward, it was taken into consideration that the two departments currently did not have such a program. 
The course scheduling in these departments is done manually on spreadsheets or whiteboards, which is a highly complex process. 
Previously, department administrators had to logistically setup physical resources like number of classrooms, time frames, and specialized equipment for courses every academic year serving as many students as possible. 
The course assignment is done based on eliciting professor teaching preferences, classroom sizes, times and labs. 

Hence, the requirements for this project included allowing users to: (1) assign professors to classes for a whole academic year, (2) assign academic courses to time slots, and (3) determine course size (i.e., number of registrants) based on expected cohort size for given semester. 
The scope of the project was limited to courses required for the software engineering undergraduate degree in the Faculty of Engineering.

\begin{figure}[]
  \centering
  \includegraphics[width=8cm]{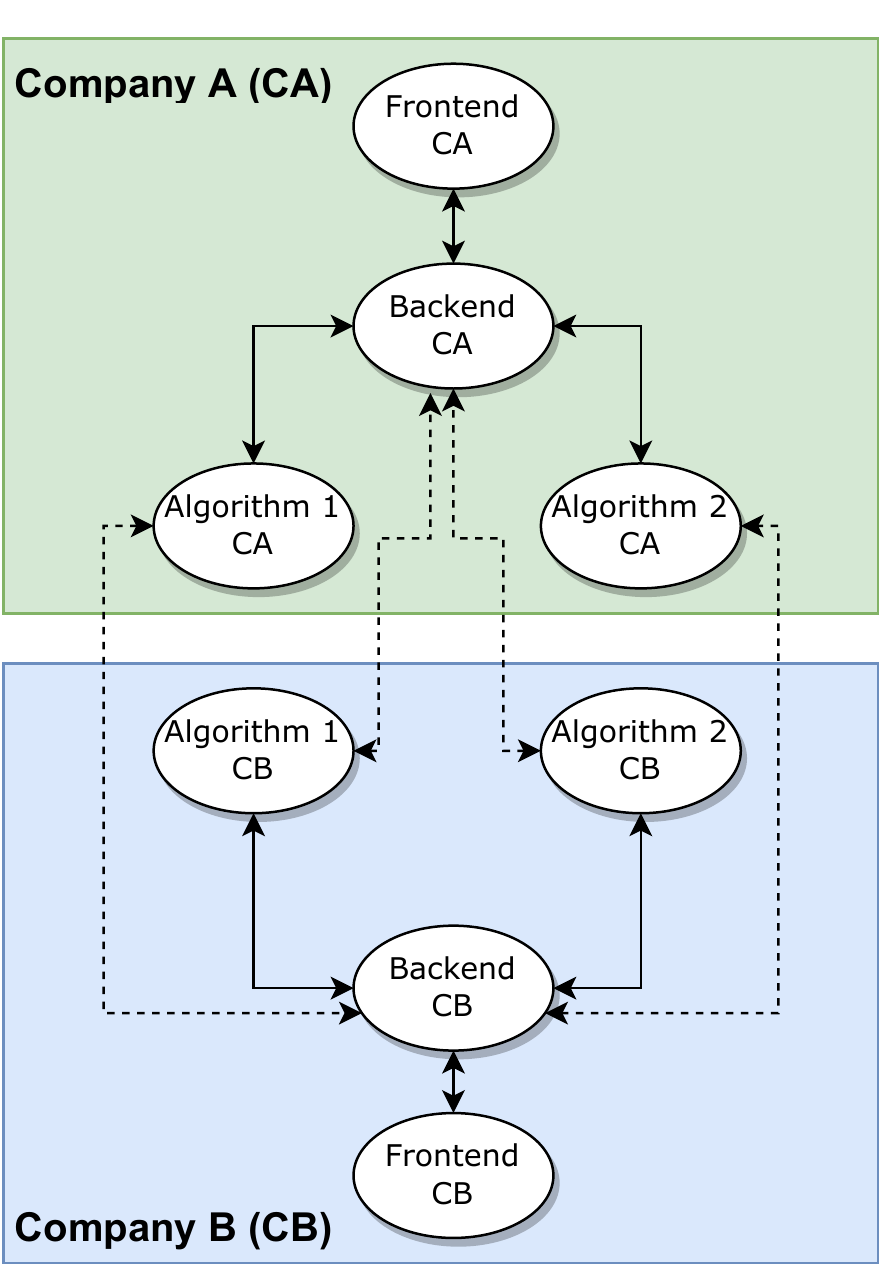}
  \caption{ Microservice Architecture For Two Companion Companies with Plug-and-Play for the Algorithms (Dotted lines represent the plug-and-play between companies)}
  \label{architecture}
\end{figure}

Although this project overall seemed straightforward, in reality the scheduling algorithms are considered to be NP-Complete problems which is a class of problem that still has no efficient solution \cite{noauthor_np-complete_nodate}.
The course scheduling was particularly complex as students would also need to consider professor sabbatical or other course leaves, as well as preference and expertise for each type of course. 
The project further required the teams or \company{companies} to incorporate a micro-services architecture as shown in Figure \ref{architecture}. 
A microservice architecture encompasses a collection of smaller applications with independent responsibility where each part makes up the large application \cite{google_noathor_microservices}. 
In this project, two companies would work under one Teaching Assistant (TA) and they would be called companion \company{companies}.
Each \company{company} would have to negotiate a set of shared services, which would allow the sub-teams to plug and play different sub-team components.
For example, frontend sub-team of Company A might want to test their interface assigning a professor to a course, but Company A’s Algorithm 1 sub-team service for that feature is not currently available. However, if the Algorithm 1 from Company B is available, Company A can use that as an alternative.

\subsection{Team Structure:}\label{team structure}
The team structure was a critical part of replicating the multi-team setting in the industry; as such, the 77 students were distributed into four \company{companies} of 19-20 students. 
To have a team distribution along aspects related to educational preparedness and industry experience, a survey was conducted at the start of the term to identify the students knowledge level.
The main criteria of selections were 1) number of experiences working on non-academic industry projects, 2) experience of coordinating between multiple teams, 3) level of technical knowledge with respect to (algorithms, frontend frameworks, database, HCI, requirements engineering) and 4) their interest in working on a particular type of sub-team.

Based on the survey responses, the students were placed in specific \company{companies}.
Each \company{company} was assigned a company-wide leader, ``scrum lord" who would be responsible for the overall \company{company} culture, meeting schedules, compiling the sprint planning, demo, and retrospective etc. 
Each sub-team in a \company{company}, further had a leader who was referred to as scrum master. 
The ``scrum lord" coordinated communication and planning with the ``scrum master'' for each sub-team.
The responsibility of scrum masters consisted of assigning and delegating the tasks within the sub-teams, leading sub-team decisions regarding features.
More experienced students took the mantle of being scrum masters.
These students frequently served implicitly as technical leads in their sub-teams and helped less experienced colleagues.
The selection of the scrum lord and scrum master was based on a team wide decision and preferences. 
A TA was responsible to support and oversee the progress of each pair of companion \company{companies} and assess them on teamwork and individual contribution. 

Following the microservices architecture in figure \ref{architecture}, each \company{company} was divided into 4 sub-teams: frontend, backend, Algorithm 1 and Algorithm 2.
Each team's responsibilities enclosed the following:
\begin{itemize}
    \item \textbf{Frontend:} Developing the user-interface (UI), client facing aspect, etc. 
    \item \textbf{Backend:} Working on databases, REST API server, GraphQL queries, standardizing data and requests between other sub-teams, etc.
    \item \textbf{Algorithm 1:} Building the main scheduler algorithm and interface with the other three sub-teams, depending on their architecture design.
    \item \textbf{Algorithm 2:} Creating a service to predict the required class sizes and class sequences so that each software engineering student can take their required courses, whether they are on-stream (i.e., on schedule in their degree) or off-stream.
\end{itemize}

\subsection{Project Timeline:} \label{projectTimelineSection}
Each of the four \company{companies} were assigned to build their own version of the university scheduler over the course of 13 weeks.
The course was divided into weekly lectures and lab work.
The first 4 four weeks were heavily focused on intensive learning on the project and relevant domain knowledge. 
Although an initial requirement document was provided to the students from the teaching team, a domain expert panel was organized during Week 3 where students could elicit additional requirements directly from the clients.
In our case, the clients were the administrators in charge of scheduling for BSEng courses.

In parallel with the project work, several class discussions on assigned readings or expert panels with guest speakers from the industry were held. 
Five industry panels were organized, and speakers from different roles, such as CEO, CTO, product manager, technical leads, and experts from their respective fields, were invited. 
The class lectures, assigned readings, and panel discussions consisted of topics like (1) Critical software engineering skills, (2) Unspoken rules of starting a career, (3) Coordination and communication, (4) Knowledge management, and (5) Culture and trust. 
During the panels, the students were expected to participate and discuss their experience working on this project to leverage the industry panelist's expertise. 
In addition, students were assigned to write blog post reflections connecting the topics 2-5 with their experiences and project.

The labs on the other hand, consisted of sprint activities such as planning, demo and retrospectives.
Week 5 to 12 were dedicated to the 4 sprints, where each \company{company} had the autonomy to decide their own deliverables, but the solution had to be complete by the sprint 4. 
During the sprints each \company{company} self-organized their own milestones and planned their deliverables.
However, all \company{companies} more or less had similar deliverables across 4 sprints which we summarize as follows: 

\textbf{Sprint 1: }
In Sprint 1, the student milestones consisted of - creating API endpoints for all components, implementing course assignments to time slots based on historical data, and setting up CI/CD (Continuous Integration and Continuous Delivery) for unit testing and integration testing. 

\textbf{Sprint 2: }
Sprint 2 milestones mainly focused on initial schedule generation for courses based on researched algorithms, implementing professor assignment to courses algorithm, manual schedule editing functionality, and full stack functionality for one happy path of each main feature.

\textbf{Sprint 3: }
By Sprint 3, the students were ready to show the complete end-to-end tests of system functionality, deployment of components to AWS/Azure/Heroku, schedule generation involving teacher preferences, and implementing user authentication. More importantly, students should have basic integration between companion companies set-up.

\textbf{Sprint 4: }
Spring 4 primarily emphasized bug fixes and validating the robustness of plug-and-play functionality with the companion company.  
The last week was dedicated to the final presentation of the projects. 
Figure \ref{coursetimeline} better demonstrates the project timeline and the activities during each week. 

\subsection{Project Deliverables}\label{deliverables}
Although the main focus of the course was developing the project to assess the learning outcome, there were few other deliverables throughout the term. 
The list of deliverables are: 
\begin{enumerate}
    \item Requirements specification document
    \item Planning, demo and retrospective for each sprint
    \item Peer review for each sprint
    \item Logbook for each sprint
    \item 4 Reflective blog posts based on required readings
    \item Project documentation and final report
    \item Final presentation
\end{enumerate}

In the last week of the capstone, each \company{company} was required to present a live demo of the solution to the client and a final presentation.
Furthermore, each \company{company} was required to write a final report comprising all the design decisions, assumptions, APIs, use cases, and onboarding guide about their solution, as documentation is a significant part of the industry. 

The students were graded on all the above mentioned deliverables. 
As a preventative measure against loafing, each student was required to complete a peer review and logbook for each sprint.
In the peer review, each student had to fill out a form where they evaluated each member of their sub-team, including their self, between the grade of A to D based on member performance and contributions to the team.
This peer review occurred after each sprint, hence, four rounds of peer reviews occurred. 
The peer reviews for each sub-team were aggregated by the teaching team to help identify potential areas where students struggled or exhibited to social loafing.
Students were given the instructions to provide at one paragraph of description for each member of the sub-team and to describe areas related to communication, completion of tasks, and participation to meetings and discussions.
Teaching team also tried to triangulate the peer review evaluations with the logbooks, where a student would log the activities and the number of hours of each activity they completed for a sprint.
Other members of the sub-team were required to anonymously provide feedback stating if they agree or disagree with a student's logbook.
These assessments were comprehensive, but needed so the teaching team could measure and validate student contribution towards the project and provide students with a sense of accountability. 
The teaching team spent a total of 130 hours each on the instructing and assessing the deliverables of the course.  

\subsection{Final Products}
By the last sprint and final presentation, each \company{company} had running versions of the program. The required features were largely all implemented by each \company{company} and the final reports provided by each \company{company} had straightforward onboarding documentation to assist with setup of the projects.
Two of the four \company{companies} (i.e., one pair of companion companies working with a TA), successfully implemented full ``plug and play" architecture. 
In other words, this pair of companies managed to achieve ``plug and play" capabilities with each of the four sub-teams. 
The frontend team of one company could interact and utilize the APIs from the backend, algorithm 1, and algorithm 2 of the other company.
Unfortunately, the two of the four \company{companies} (i.e., the other pair of companion companies with with the other TA), was unable to get ``plug and play" for all the subteams working.
The two \company{companies} achieved ``plug and play" for the Algorithm 1 and Algorithm 2 subteams, but did not establish ``plug and play" for the backend.


\section{Lessons Learned}
In this section, we describe the \textit{lessons learned} from the first iteration of our revamped capstone course from the perspective of both the students and the teaching team. 
To acquire a broad understanding of the student perspective on the course, we collected reflections from approximately 70 students.
Students were required to write individual reflections as part of their final deliverable.  
They needed to provide a critical analysis of their experience in the course discussing accomplishments and lessons learned, which include positive and negative perspectives.
Students were asked to provide grounding to the readings, and conveying learning and personal growth through the experiences in the project. 
The reflections were instrumental for understanding the course from the students' perspective.
We also provide \textit{recommendations} for improvements for subsequent offerings of the course in our university as well as for other universities and programs who may be interested.
In our new capstone course, students were presented with a unique blend of autonomy and course experience where they worked in massive software \company{company} setting requiring them to coordinate and communicate amongst multiple teams.
Our students experienced challenges along the way, but significant benefits to their personal education were realized in this experience.


\subsection{Student Perspective}

{\textbf{Lesson 1:  \textit{Team culture and trust are important pillars for large-scale, multi-team student projects to succeed, though culture and trust take time to build and conflicts may occur.}}} 
We noticed that different \company{companies} formed different cultures, however the difference was mostly noticeable at \company{company} sub-level.
Many students questioned whether self-organizing the large \company{companies} in three months was possible.
As one student said, \q{How can a group of twenty individuals, many of whom strangers, organize themselves to build an automated scheduling tool in three months? This is the question I asked myself at the beginning of class once the project was revealed... At the end of the course, our company [looked] back at the hurdles we conquered together and be proud of the product we produced.}

Anticipating that culture and trust would impact team success, one of our industry panels and set of required readings were about culture and trust. 
In one \company{company}, the students were able to bond from the beginning.
\q{A lot of trust was built very quickly in part by just searching for someone to follow in the initial chaos of the groups. I do feel that [redacted] was an amazing scrum master and I think the style and energy really affected how our company functioned as a whole...
I think [redacted] helped create in general a more friendly environment that reduced a lot of issues.}

Trust between \company{company} members grew as friendships emerged in the foreground of the projects.
Some \company{companies} built culture and trust through social gatherings and completing work together in labs.
\q{Our company had a lot of fun with the class, doing themes [i.e., dressing up differently] for each demo, meeting at [campus pub] after the labs, and generally having fun while working in the labs.}
For \company{companies} that did not establish a culture of social events, some students expressed regret. 
\q{Our company did no social events apart from the Karaoke Night [i.e., course wide social event organized by teaching team], in hindsight, it would have benefited us greatly to have such kind of events more often to help people talk to each other in informal settings.}

For such multi-team setting, building a culture where everyone has a sense of belonging helps with overall achievement of the project.
One student explained, \q{I was impressed by the collectivist values that my company had. Once we began setting goals at a teamwide scope, each sub-team worked hard to achieve those goals. We were much more
concerned about the group as a whole succeeding rather than having individual success.}

However, some \company{companies} had a more rigid structure, particularly, from the scrum lord. 
Regretfully, this type of culture has cascading negative effects, which can create an environment where blame is assigned as opposed to objectively resolving problems. 
\q{
[Redacted] was micro-managing and acting condescending to the company members. While I believe [redacted] was able to overcome their issues they did have a boiling point of tension when [redacted] called out [redacted] for sprint issues during a retro instead [of] using blameless post-mortem to address company failings.}

Since students worked closely with other students in the same sub-team on modules and features for the same microservice, we found issues typically pertained to between \company{company} or inter sub-team situations.
In one aspect, a student dealing with the plug and play microservices aspect described their observation of  individuals pushing their ideas through.
\q{I [worked] closely with this individual [from other company] on the plug and play operations and format of our algorithm. Because this individual was very outspoken, and pushed their own ideas, it ended up having a significant effect on both teams architecture and how data was transferred from all components... One person being slightly about their own designed decisions at the start of the project ended up having such significant design implications for 40 other people... This was a great example of “The squeaky wheel gets the grease” in a professional context.}


Another student described how members of their \company{company} lacked confidence in the \company{company}. \q{Our scrum master also seemed to lack trust in the company. [Redacted] would frequently interrupt other sub-teams to critique their systems, but not give insight on how to improve their system. 
I think our scrum master went beyond the scope of effective scrum and into the realm of micromanagement. 
In the future, I think it would be best to let the entire class know that a scrum master of a company is there to facilitate productive practices and is not the designated CTO or CEO.}
Putting individuals in ``power" can result in situations where those individuals may fail to consider trust and teamwork. Previous work suggests that giving individuals power may open a door to abuse power. Since power can be acquired through diverse means \cite{vredenburgh1998hierarchical}, we share suggestions to select scrum masters strategically in Section \ref{teachin-team-perspective}.
Though fortunately for our course, this situation was isolated, we anticipate future educators may encounter such issues.


Meeting deadlines and struggling with time constraints created other situations of conflict and hindrances of trust for the multi-team coordination. 
Nearing a major deadline, a student recounted, \q{While I offered my help a few times something [redacted] said, ``I really appreciate it but at this stage, it’ll take too long to get you up to speed and we just can’t right now."}


Building trust and a culture where members are supported was an important dimension for project success. 
One student summarizes this conclusion, \q{
The social relationships and trust that you build within team takes a lot of time and effort to grow, but by promoting a culture that incentives these unspoken rules [i.e., thinking like an owner and taking ownership and agency over your responsibilities, as well as showing that you want to learn and help others to achieve the team’s goals], I believe that in the end it will be easier to discuss ideas, make compromises, and make decisions with people feeling like they had a say in the final decision.
}

\begin{mybox}
\textbf{{Recommendations:}}
\begin{itemize}    
    \item Teaching team may provide ice breaker activities or strategies for building trust and relationships 
    \item Students should be encouraged to engage in team bonding activities. The course wide social event organized by the teaching team was instrumental for building trust and friendships for companies, particularly those companies that did not have prior social gatherings of their own.
    \item Teaching team should consider adding an interview step for prospective scrum masters. 
    \item Conflicts likely will occur, particular at the inter sub-team and inter company level. Teaching team should be prepared in advance for intervening in resolving conflicts. 
\end{itemize}

\end{mybox}

{\textbf{Lesson 2: \textit{Full autonomy to self-organize was given to students so they could experience working in a multi-team environment, but additional support may be necessary to guide the large \company{companies}}.}}
The students received full autonomy for how they wanted to organize their \company{companies} and sub-teams.
Most students thrived with the full autonomy and lauded the self-learning aspect about the course.
As one student explained, \q{Crucial difference in this course that makes it stand out from all other course offerings at the university is the immense shift towards self-learning and independence from a teaching/supporting team. The amount of work needed quickly compounded from one sprint to the next - which meant I needed to explore new approaches to learning and more importantly collaborating with others.}


\company{Companies} had the autonomy to decide how they organize their repository and expectations for documentation.
However, the autonomy for multi-team coordination did not always lead to desirable outcomes \q{
When I look at other subteams' repos there was no unified structure or documentation method
on a company wide level. Every subteam was operating in their own ways with no immediate consideration for integration. Most of the knowledge was transferred by mouth or slack threads and not all meeting notes were shared with the rest of the company. Some subteams had no documentation or tickets... In sprint 3 we appointed regular subteam members for integration to introduce more links between subteams and it proved to be beneficial. It lessened the responsibility load on team leads.
}

Similarly, some students from \company{companies} expressed frustration that autonomy for sub-teams allowed them to self assign deadlines, which may not have aligned with \company{company} goals and this hurt the trust in the \company{company}.
For example, \q{For the last two demos we agreed on timelines for the algorithms to finish all of their work so the backend could integrate with them, and then deadlines for the backend to finish that integration, so the frontend could do a full end to end test before the demo. For both the last demos, one of both of the algorithm teams did not finish all their work and were not deployed when they said they would be. This negatively affected the backend team as we now had way less time to test that integration and get it working properly. This definitely affected how I viewed some other sub teams and I felt negatively impacted our ability to come together into one cohesive company.}
The issues with autonomy in the multi-team \company{company} were often linked with aspects such as knowledge management and communication.
Fortunately, through the multi-team experience, our students were able to improve on these soft skills during the project.


The autonomy also facilitated students self assigning roles in their respective sub-teams.
One student desired to be a technical lead and felt that being a scrum master provided crucial experience in expediting this goal. \q{[I am] graduating this term and aspire to be a technical lead at some point, I believe the structure of this Capstone project gave me valuable experience that I will need down the road... I can say that experience in it itself was an accomplishment and one that I will continue to learn from as I endure it in a professional workplace.}
Another student described how they joined a sub-team to develop specific technical skills. \q{My goal was to jump on a team I did not have much experience with before. I joined [redacted] team for the project... learned tools I had never used before... cooperated with members that had more experience
}

Despite the potential challenges of autonomy in the multi-team project, taking charge and having the means to dictate the direction of the project created a sense of pride and self-ownership.
As one student described, \q{Over the course of the last 3 months, being sorted into sub-teams and forced to create our own tickets and come up with the approach we thought best, I felt real responsibility and ownership over this project.}
In contrast, some students felt that having more structured guidance as opposed to the autonomy and mostly self-learning would have been beneficial to the student experience. This may be dependent on previous experience, as autonomy is typically more motivating when an individual possesses more competencies \cite{noll2017motivation}.


\begin{mybox}
\textbf{{Recommendations:}}
\begin{itemize}

    \item Providing the sense of ownership and autonomy over the project may help student motivation.
    \item For autonomy to work well in multi-team projects, teaching team should be ready to provide guidance and structure during the semester, especially for students who may prefer more structure.
\end{itemize}

\end{mybox}

{\textbf{Lesson 3: \textit{While technical abilities are important for students before they leave school, soft skills such as communication and coordination are paramount to the success of any multi-team project.}}}
In the capstone, students are expected to demonstrate their technical knowledge developed over 4 plus years.
However, it was also an opportunity for many students to learn new languages (e.g., Go, JavaScript), databases (e.g. MySQL, PostgresSQL), APIs, (e.g. REST, GraphQL), algorithms (e.g. branch and bound, reinforcement learning, genetic algorithm, linear regression), and tools (e.g. OpenAPI, Zenhub, Prisma).
In contrast, we dedicated class discussions, required readings, and industry panels about soft skills that we expected to immensely affect the coordination and teamwork of the large \company{companies}. 
We specifically allocated time to cover topics such as the Unspoken Rules for a young professional entering the work force \cite{ng_unspoken_2021}, coordination and communication, and knowledge management. 

Since very few students had experience working with multiple teams, \textbf{communication} between sub-teams and \company{companies} was a struggle, especially early on. 
Students slowly developed more effective communication skills transcending each stage of the project.
The effects of Conway's Law \cite{conway1968committees} on the multi-team coordination was quite noticeable as one student described,
\q{Conway’s law has been in effect the entire time, but in reverse...
From the start there was no communication between subteams, so the individual pieces of code did not communicate either. Later on there was communication between subteams, but not companies, so the individual products worked but not plug and play. We built up to our final communication network over the term, and it has a striking resemblance to the architecture of the final product.}
Moreover, under-communication was described as a common problem that they learned to avoid, \q{My biggest takeaway [is that] coordination and communication needs to be a top priority early on, and it’s difficult to accidentally over-communicate in a project. We naturally leaned towards not communicating during times of uncertainty when these were the times we needed the most communication.}


\textbf{Knowledge management} was another soft skill that students developed throughout the semester, despite many starting with limited experience. 
\q{[We] had a lack of skill and experience with knowledge management. [It was hard] especially with teams meeting both in person and in a distributed fashion - multiple times per week within their sub-teams. We often struggled with large knowledge gaps, overlapping/repeated work, unclear/different understanding of requirements, and general confusion between groups in the subteams, the subteams themselves, and our company as a whole, especially when working on plug and play.}

As one student described, documentation and communication were important for establishing knowledge management. \q{I worked with knowledge management throughout the semester, I gained a newfound respect for proper documentation and having communication channels between the subgroups.}
Unfortunately, documentation was a constant challenge as some teams made trade-offs with respect to good documentation.
\q{Although some of us started out with setting standards for documentation and branching structures, when it comes to integration hell it is so easy to deprioritize quality for hacks. Hacks are supposed to be temporary but end up as code debt as the system grows and there’s no incentive to improve due to the next deliverable.}
Maintaining documentation standards was tenuous due to the purported lack of incentive.
However, we tried to mitigate this issue through the project documentation and final report assignment.

Some \company{companies} organized themselves into sub-teams right away and completed most work from the perspective of the sub-team.
The fragmentation led to coordination and integration problems. 
\q{Fracturing was done so that we could each focus on some subset of requirements, but [ended up causing] integration issues during development. Each subteam room [created] their own ``mental image" of the end product, [but] it is crucial to come together and combine the mental models before creating the final list of requirements or defining any specifications. This ``combination" step was done very informally by our company. This resulted in a ``Frankenstein" document, full of contradictions and even logical errors.}
Knowledge management is not just about documentation, it is also about tacit knowledge and communicating those to the team. 
As one student reflects, \q{I learned that tacit knowledge is at the foundation of communication.}

To ensure that the microservices and plug and play architecture would succeed, each paired \company{companies} had to conduct coordination meetings, which include 8 sub-teams between a pair. 
For plug and play, some companies initially started with massive meetings attended by all members, but often resulted in ineffective meetings.
One student described this experience, \q{Up until this project, I had only truly experienced at most 10 developers in a team even with my coops. So, to witness the absolute chaos of 20-40 people trying to coordinate on decisions was very interesting to see. One of the ideas from the readings that came true within the project was that the needs of the project evolved over time. Initially, the individual teams (backend, frontend, algorithm1, algorithm2) were able to work in isolation without any real coordination with each other. However, as integration became a more prevalent issue, the need for a dedicated deployment and integration team arose.}

However, this does not mean team members should be completely abstracted from the design or API meetings.
Students describe losing the bigger picture if not updated on decisions. 
One student stated, \q{[Scrum masters in a company] meetings turned out to be less effective than just having a company wide standup where each scrum [master] updated the rest of the subteams with what their team had been doing. This allowed for the whole team to be kept in the loop. Our initial scrum leader meetings seemed good, until other members of the team expressed they felt like they didn't know what had been decided. By taking this feedback and shifting to team-wide meetings we improved coordination.}
Likewise, a student added, \q{I came to the realization that it wasn’t feasible for me to technically understand everything going on within our team of 20, but if I had a good grasp on what my team was responsible for and how what we were doing was affecting the other teams I could effectively contribute without drawbacks.}


In theory, technical and soft skills are constantly developed through courses and work experience.
However, one student summarized the importance of this course, \q{Unfortunately, there is no quality control on what is learned during your co-op [internship]. I think this course solves that problem. It allows students to experience the challenges of collaborating in large teams and teaches methods for overcoming said challenges, which is important in preparing them for what lies ahead after leaving university. Sure, knowledge management, communication, and organization are all skills that are developed throughout the software engineering degree, but not at this scale.}

\begin{mybox}
\textbf{{Recommendations:}}
\begin{itemize}
    \item Students develop technical skills as part of working on the project, but teaching teams should anticipate and plan ahead of time some of the soft skills that students should learn from the course.
    \item In a large-scale multi-team project, \textbf{knowledge management}, \textbf{coordination}, and \textbf{communication} are critical success factors to the project. Teaching team can take advantage of assigned readings, blog post reflections, industry expert panels, and in class instruction to strengthen these skills. 

\end{itemize}

\end{mybox}


\subsection{Teaching Team Perspective} \label{teachin-team-perspective}

To understand which successful aspects of the course should be retained, and which should be altered for future offering of the course, the teaching team came together to reflect on their experience. 
We report the following lessons from the teaching team's perspective for future instructors to consider if they offer a similar course.

{\textbf{Lesson 1: \textit{Team size should vary with the scope of the project.}}}
Amongst the teaching team there were some mixed opinions on the optimal student-team size. 
In particular, the TAs were concerned regarding quieter students receiving significantly less interaction amongst multiple teams as they were often overlooked due to stronger personalities. 
One TA says \q{This is the problem with such large teams...the strong personalities, were the only ones that really interacted with me at any significant level...and that was a bit of a problem, because I only saw a snapshot of like, who the students were.}  
Previous work by Horwitz  \cite{horwitz2005compositional} suggests that team performance on complex tasks increases in tandem with team size, but only up until a certain threshold. 
That is, adding members to a team is beneficial because of easier division of labour, and exposure to more perspectives to solve a complex task. 
However, in adding members there comes a point where there are more negatives than positives. 
Too many members on one problem can create conflict especially in simpler tasks \cite{horwitz2005compositional} and makes the team prone to a phenomenon known as \textit{social loafing}, where team members begin to put in less effort than they would exert individually due to the perception of being able to rely on others \cite{mao2016experimental}.

Although the other TA describes the issue of team size with less severity, they still acknowledge that there appears to be a \q{core team} from each company interacting with the teaching team as opposed to direct interaction with all students. 
This TA mentions that \q{it definitely felt like there were a lot of passengers in the bus, as opposed to drivers in the bus}, suggesting that social loafing was observed, and that the team size may have been slightly too large for the scope of this project. 
Since the ability work in large teams while coordinating amongst multiple teams is both an important skill for graduating software engineering students, and a learning outcome of this course, we speculate that the scope of the project may not have been broad enough for our intended team size. 
That is, the project should not be doable without each team member contributing valuably. 

In large scale, multi-team development, \company{companies} are expected to be self-managed, however structure is provided through management, milestones, and company goals in industry. 
To compensate in the academic setting, we also recommend adding structure into such a capstone course so that work within teams is properly distributed. 

Although, based on peer review process mentioned in Section \ref{deliverables}, individual grades were adjusted to have a fair grading system.
However, it is perceived that more structured work distribution methods would ensure that all students contribute to the project significantly, and prevent burnout for those students who would otherwise suffer for their peers' minimal efforts.
Overall, more forethought is required in deciding on the team size so students can have equal opportunities the multi-team collaboration. 


\begin{mybox}
\textbf{{Recommendations:}}
\begin{itemize}
  
    \item To teach students how to work in multi teams, the scope of the project should be large enough such that there is capacity for all students to contribute significantly.
    \item Educators should include a component for peer evaluation of student contributions. 
    
\end{itemize}

\end{mybox}

\noindent
\textit{\textbf{Lesson 2: Ample thought is needed in scrum master selection.}}
The scrum master in agile development processes has an abundance of responsibilities, such as ensuring that the team is functioning effectively, and helping the team overcome obstacles \cite{shastri2021spearheading}. 
As a result, the personality of this leader has a direct effect on the team members and their ability to work well. Consequent to a capstone course, this means that the student selected for the scrum master position will have an impact on how their teammates perform in the course. 
The teaching team identified from observing the project teams that the role of the scrum master is important \q{when [they] are delegating [to their] fellow classmates in a large group project with arguably high stakes in terms of a course group project}. 
Despite all of this, there is a significant knowledge gap in what tangible responsibilities scrum masters actually have. 

The teaching team observed the need to clarify the scrum master's responsibilities and also the importance that this role impacted teams.
The instructors learnt that, \q{more focus on roles within the company and clarity around [scrum masters/lords] responsibilities would also help. It seemed to [the students] that certain scrum masters/lords were more knowledgeable/experienced than others and this unfairly effected the entire team}. 
It appears that it is critical for students, even those not elected as scrum masters, to properly understand the role a scrum master should play within the team. 
The relationship between scrum masters and their teams is further complicated in the academic context due to the inexperience of students acting as scrum lords. 

For instructors, these experiences suggest that the selection of scrum masters is non-trivial. While inexperience is rather inevitable with students, looking for certain personality traits like openness and agreeableness may be important preconditions for fulfilling this leadership role successfully \cite{kornor2004personality}. 

\begin{mybox}
\textbf{{Recommendations:}}
\begin{itemize}
    \item Implement a screening/interview process for students interested in assuming the scrum master/lord position to ensure other students are exposed to adequate leadership. 
    \item Allow students to rotate roles during the semester so that multiple students can gain leadership experience.
\end{itemize}

\end{mybox}

\noindent
\textit{\textbf{Lesson 3: Students require support in integrating their technical skills in a capstone project.}}
The teaching team recognized the need for an increased emphasis on technical skills in this course. 
The material taught in this course, aside from the actual project work, focused largely on soft skills such as company culture as by the final capstone course, we expected students to be proficient in foundational software engineering technical skills. 
Furthermore, companies are placing an increased emphasis on soft skill competencies when hiring new employees \cite{medlin2001students}, thus, this emphasis on soft skills was also done with students' best intentions in mind. 
As an example of soft skill training, the panel sessions in this course were comprised entirely of soft skill topics facilitated by industry experts such as CEOs, CTOs, and product managers. 
While the students engaged with these soft skill sessions, the instructors noticed that students \q{learned why culture is so crucial for any company being successful}. 
While the emphasis on soft skill development was perceived as valuable for many students, it seems that the emphasis on technical competency was not prominent enough. 
Instructors observed that in order to enhance the student experience, \q{more review, and guidance on how to integrate requirements, evolution, HCI, etc. would help.} 


One aspect students struggled with was managing the volatile nature of requirements change.
Throughout the course, the instructors found the students discussing that
\q{[they thought] one aspect that was difficult to adapt to was the ambiguity of the requirements on top of the speed that they changed at... This did, however, teach [them] to appreciate documented requirements and the ability to change [their] path towards requirements that were not agreed upon initially – both very relevant to work in industry.}
We also recommend that within technical training, an emphasis be placed on the \textit{combination} of skills as opposed to individual skills. 
After a four year software engineering degree, students should have obtained a variety of skills, but the technical difficulties lie in combining them. 
The teaching team observed the importance of this course for helping students combined a variety of technical skills and was best corroborated by a student, \q{this course ended up being a positive challenge to learn about processes which otherwise I would not have known about from a practical standpoint. For example, the importance of a great API specification, and the existence of one source of knowledge, particularly for inter company communication where a lot of communication channels would be required}. 

\begin{mybox}
\textbf{{Recommendations:}}
\begin{itemize}
    \item Since many students learn a variety of technical skills in isolation, technical instruction may focus on the integration of these skills rather than each skill separately.
\end{itemize}

\end{mybox}

\subsection{Threats to Experience}
Even though this manuscript was an experience report, we acknowledge several threats that could have impacted the reflections on this experience. 
The first threat to our experience was that assessing the learning of the students could have been limited as we relied on primary sources of data: student personal reflections at the end of the semester and teaching team observations. 
We trust that students were honest about the skills that they learned in the course and to mitigate against bias in the reflections, we explained to students that they could opt out from the research. 
The reflections were meant to provide deeper insights about the course, and the students were encouraged to critique the course. 

The second threat to our experience stemmed from the occurrence of social loafing in some of the sub-teams. 
Social loafing is a common phenomenon that occurs in project teams and we tried to reduce the potential of social loafing in these projects through the introduction of peer reviews and logbooks. 
However, social loafing can still occur undetected if members in a sub-team are complacent in reporting social loafing when they observe it. 
Since the course is fast paced with only four sprints, it is also possible that by the time students can no longer tolerate social loafing and starts reporting it in the peer reviews and logbooks, the course is almost complete. 

The third threat to our experience was that while the student cohort is quite experienced with at least three to four internship experiences, many lack management or leadership experience. 
As a result, while the teaching team did its best to mimic a real world large-scale, multi-team environment, students when they go into industry would most likely get placed in teams with experienced scrum masters and managers who have more experience delegating work.

\section{Conclusion and Future Work}
Capstone courses in software engineering and computer science programs are vital to a student's education as it is the final stepping stone in their academic career before graduation.
In this paper, we introduce our new capstone course for students enrolled in the Bachelor of Software Engineering program at the University of Victoria. 
Our capstone course places students in large \company{companies} of 19-20 students and were tasked to build a course scheduling system following microservices architecture and common APIs to enable plug-and-play between \company{companies}.
The project requires a student \company{company} to develop an application where professors in software engineering and computer science department can provide teaching preferences and administrators can generate course schedules over a semester. 

We observed students benefiting as they enhanced not only their technical skills working in a large-scale, multi-team project, but also improved their soft skills such as communication and coordination.
Combining their prior software experience and knowledge for the project, students reflected that this course was more impactful and significant to their education than any prior courses in their degree.

In future semesters we will run this course again with a similar structure (i.e., multiple sub-teams), but with different types of projects and observe if that impacts the lessons learned.
Further, we may experiment with slightly smaller \company{companies}, such as 14-16 students, as opposed to 19-20 students, which may assist in reducing social loafing. 
We report several recommendations and lessons learned for future iterations that other universities and programs may find fruitful for launching their own capstones. 
Working in a large-scale software project significantly develops a student's abilities such as communication and knowledge management to work in multi-team environments, and trust and culture in a large project team can take time to build and teaching team should support teams building trust.

\section{Acknowledgements}
We would like to thank all the students in the SENG499 Capstone Course for their feedback and insights to help improve the course. 
We also want to extend our gratitude toward Amanda Dash for her help running the course and sharing her experiences from the course. 

\bibliographystyle{IEEEtran}
\bibliography{main}

\end{document}